\documentclass[usenatbib]{mn2e}

\usepackage{graphics}
\usepackage{amssymb}
\usepackage{epsf}

\newcommand{\sini}{\ensuremath{\sin i}}

\newcommand{\ve}{\ensuremath{v_{\rm e}}}
\newcommand{\vesini}{\ensuremath{\ve\,\sini}}
\newcommand{\vsini}{\ensuremath{v\,\sini}}
\newcommand{\vcrit}{\ensuremath{v_{\rm c}}}
\newcommand{\geff}{\ensuremath{g_{\rm eff}}}
\newcommand{\teff}{\ensuremath{T_{\rm eff}}}
\newcommand{\logg}{\ensuremath{\log{g}}}

\newcommand{\Msun}{\ensuremath{{\rm M}_{\odot}}}
\newcommand{\Lsun}{\ensuremath{{\rm L}_{\odot}}}
\newcommand{\Rsun}{\ensuremath{{\rm R}_{\odot}}}

\newcommand{\Mstar}{\ensuremath{{M}_{*}}}
\newcommand{\Lstar}{\ensuremath{{L}_{*}}}

\newcommand{\Rpole}{\ensuremath{R_{\rm p}}}
\newcommand{\Req}{\ensuremath{R_{\rm e}}}

\newcommand{\Tpole}{\ensuremath{T_{\rm p}}}

\newcommand{\grpole}{\ensuremath{g_{\rm p}}}

\newcommand{\kms}{\ensuremath{{\rm km}\,{\rm s}^{-1}}}

\newcommand{\wid}{\ensuremath{\Delta}}
\newcommand{\widu}{\ensuremath{\wid_{\rm u}}}
\newcommand{\widc}{\ensuremath{\wid_{\rm c}}}

\newcommand{\vdef}{\ensuremath{\delta V}}

\newcommand{\bruce}{\textsc{bruce}}
\newcommand{\rotcolor}{\textsc{rotcolor}}
\newcommand{\tlusty}{\textsc{tlusty}}
\newcommand{\synspec}{\textsc{synspec}}

\newcommand{\heone}{He\,\textsc{i}}
\newcommand{\mgtwo}{Mg\,\textsc{ii}}

\newcommand{\eg}{e.g.}
\newcommand{\ie}{i.e.}
\newcommand{\cf}{cf.}
\newcommand{\etc}{etc.}

\title[Be-star rotation:  how close to critical?]
      {Be-star rotation:  how close to critical?}
\author[R. H. D. Townsend et al.]
       {R. H. D. Townsend$^{1, 2}$\thanks{rhdt@bartol.udel.edu}, 
        S. P. Owocki$^{2,1}$ \& I. D. Howarth$^{1}$\\
        $^{1}$ Department of Physics \& Astronomy, 
        University College London, 
        Gower Street, London WC1E 6BT, UK\\
        $^{2}$ Bartol Research Institute,
        University of Delaware,
        Newark, DE 19716, USA}
\date{%
Received: .................................... 
Accepted: ....................................
}

\pagerange{\pageref{firstpage}--\pageref{lastpage}}
\pubyear{2003}

\begin{document}


\maketitle

\label{firstpage}

\begin{abstract}
We argue that, in general, observational studies of Be-star rotation
have paid insufficient attention to the effects of equatorial gravity
darkening. We present new line-profile calculations that emphasize the
insensitivity of line width to rotation for fast rotators. Coupled
with a critical review of observational procedures, these calculations
suggest that the observational parameter \vsini\ may systematically
underestimate the true projected equatorial rotation velocity,
\vesini, by some tens of per cent for rapid rotators.  A crucial
implication of this work is that Be stars may be rotating much closer
to their critical velocities than is generally supposed, bringing a
range of new processes into contention for the elusive physical
mechanism responsible for the circumstellar disk thought to be central
to the Be phenomenon.
\end{abstract}



\section{Introduction} \label{sec:introduction}

Be stars are near-main-sequence, B-type stars that have been observed
to show Balmer-line emission.  It is approaching a century and a half
since the discovery of the first Be star \citep{Sec1867}, and in
recent decades compelling empirical evidence has associated the
emission-line episodes with the formation of quasi-Keplerian,
equatorial disks \citep[\eg][]{Dac1986,Han1996}.  However, there is
still no general agreement on the underlying physical mechanisms
responsible for disk formation \citep[][]{SmiHenFab2000}.

A phenomenologically crucial characteristic of Be stars is that they
rotate rapidly -- more rapidly than any other class of nondegenerate
star. Virtually as soon as this was recognized, it was hypothesized
that the equatorial rotation velocities, \ve, may be sufficiently
close to the critical velocity, \vcrit, for material easily to `leak'
into a disk (\citealt{Str1931a}; here \vcrit\ is the velocity at which
centrifugal forces balance Newtonian gravity at the equator, so that
the effective gravity, \geff, vanishes).  Slettebak and others later
scrutinized this hypothesis quantitatively \citep[e.g.,][]{Sle1949,
Sle1966a, Sle1992}, establishing the current canonical view that Be
stars' rotation is actually significantly {\em sub}critical, with
$\ve/\vcrit \simeq 0.7$--0.8 \citep[e.g.,][]{Por1996,Cha2001}.

If rotation does play a direct, causal role in the Be phenomenon, then
it is tempting to speculate that it does so by reducing the effective
equatorial gravity to an extent that allows `weak' processes to move
material into orbit easily \citep{Owo2003}. However, if this is to be
a promising avenue of exploration, then rotation must be much closer
to critical than is generally supposed at present: at $\ve/\vcrit
\simeq 0.7$ the equatorial gravity is reduced by only about a factor
2, and it requires $\ve/\vcrit \simeq 0.95$ to reduce \geff\ by an
order of magnitude. Alternatively, to launch material into orbit
ballistically at $\ve/\vcrit = 0.7$ requires velocities in excess of
$100$~\kms; launch velocities commensurate with the sound speed (at
which the aforementioned weak processes typically operate) are
achieved only when a rotation rate $\ve/\vcrit \simeq 0.95$ is
reached. Similar arguments may be brought to bear on the angular
momentum budget needed to achieve orbit.

In the present paper we review the bases for determining the true
projected equatorial rotation velocity, \vesini, and the relationship
to its observational counterpart \vsini\ (where $i$ is the angle
between the line of sight and the rotation axis).  We argue that
equatorial gravity darkening may plausibly have led to rotational
velocities of Be stars being systematically underestimated, and that
Struve's original hypothesis of near-critical rotation in Be stars
therefore has not been ruled out by existing observational studies.

\section{Preamble}

For a centrally condensed star, the critical velocity of rotation
is given by
\begin{equation} \label{eqn:vcrit}
\vcrit = \sqrt{G\Mstar/\Req} = \sqrt{2G\Mstar/3\Rpole},
\end{equation}
where `e' and `p' subscripts are used throughout to denote equatorial
and polar values, respectively (throughout this paper we assume
uniform angular velocity at the stellar surface). We note that, even
if it were possible to measure the equatorial rotation velocity with
complete accuracy, these expressions imply considerable observational
uncertainty in establishing \vcrit.  In particular, it is customary to
adopt masses and radii from look-up tables, on the basis of spectral
type; and yet not only is the spectral type affected by rotation, in
an aspect-dependent manner \citep[\eg][]{Sle1980}, but the physical
relationship between spectral type, mass, and radius is also modified
\citep{Mae2000}.  Thus systematic errors in \ve/\vcrit\ may easily
arise from this source alone.

The primary focus here, however, is on the determination of \vesini.
Almost all published surveys of rotational velocities in Be stars rely
on one of two approaches.  The first is to infer \vesini\ values from,
typically, line width at half depth, calibrating the results through
the standard-star system established by \citet{Sle1975} or its
extension to rapid rotators \citep{Sle1982}.  The second is to compare
observations with `spun-up' versions of narrow-line stars\footnote{Not
necessarily corresponding to slowly-rotating stars} or
model-atmosphere flux spectra, typically using a convolution approach
\citep[\cf][]{Gra1992}. We emphasize that the latter approach is
unphysical: the convolution between intrinsic line profile and Doppler
broadening function assumes the two to be independent, which is
clearly not the case when the stellar surface parameters (\ie, \teff\
\& \geff) vary with latitude as a result of gravity darkening.

Both techniques imply the assumption of a single-valued,
(approximately) linear relationship between line width and \vesini. As
we demonstrate below, the true situation is significantly more
complicated.  In particular, for a rigidly rotating star with
radiative envelope that follows the \citet{vonZei1924} form for
gravity darkening,
\begin{equation} \label{eqn:zeipel}
\teff \propto \geff^{0.25},
\end{equation}
the contribution of the cooler (and hence fainter) equatorial regions
to the observed spectrum may be reduced to the point where they are
effectively invisible.  A necessary consequence of such behaviour is
that \vsini\ values determined without proper attention to gravity
darkening will systematically underestimate true projected equatorial
velocities. This point is developed in the following sections, through
the use of illustrative calculations and by a critical review of
observational studies.

\section{Models} \label{sec:models}

In Table~\ref{tab:params} we introduce parameters for thirteen stellar
models, representative of the spectral types\footnote{We stress that
here, and throughout, our use of the term `spectral type' in reference
to our models is little more than a notational convenience: although,
for instance, a non-rotating star with the same parameters as our B0
model will exhibit a spectrum loosely resembling a star with MK type
B0, it should always be remembered that the MK scheme
\citep[\eg,][]{MorKen1973} is \emph{two-dimensional}, and variations
in a third parameter, such as rotation, negate any direct
correspondence with physical parameters.}  spanning the B0-B9
interval. The mass and polar radius of the models in
Table~\ref{tab:params} are taken from \citet{Por1996}, with the
critical velocity evaluated using equation~(\ref{eqn:vcrit}). To
obtain the accompanying luminosity data we use the Warsaw--New Jersey
code\footnote{Details of this code are given by \citet{DziPam1993} and
\citet{Dzi1993}, the only significant difference in the present work
being the adoption of more-recent \textsc{opal} tabulations for
opacity \citep{IglRog1996} and equation of state \citep{Rog1996}.} to
calculate evolutionary tracks at the masses tabulated by Porter, 
determining \Lstar\ from the
epoch at which the radius coincides with Porter's values.

For each of the thirteen stellar models, we construct two sequences,
differing only in the treatment of the centrifugal force due to
rotation. For the `complete' set of models, the effects of this force
are fully accounted for: the surface geometry is that of a
gravitational equipotential within the Roche approximation, and the
surface temperature distribution follows the \citet{vonZei1924}
gravity-darkening law appropriate to radiative envelopes (\cf\
equation~\ref{eqn:zeipel}). The luminosity \Lstar\ is preserved under
changing model geometry by adjusting the polar effective temperature
in accordance with the relation
\begin{equation} \label{eqn:tpole}
\Tpole = \left(\frac{\grpole \Lstar}{\sigma \Sigma_{1}}\right)^{0.25};
\end{equation}
here, $\sigma$ is the Stefan-Boltzmann constant, and $\Sigma_{1}$ is
the surface-area weighted gravity of the rotating star
\citep[\eg,][]{Cra1996}.

In contrast, for the `uniform' set of models the centrifugal force is
neglected, so that the star remains spherical in shape, and exhibits a
uniform surface temperature and gravity; these models mimic the
traditional methodology developed by \citet{Car1928} and
\citet{ShaStr1929}, which is the basis for the convolution approach
described by \citet{Gra1992} and used by many authors (e.g.,
\citealt{How1997, Cha2001}).

\begin{table}
\caption{Fundamental parameters for the B-type models considered
herein.}
\label{tab:params}
\begin{tabular}{@{}cccccccc} 
Sp.      & \Mstar  & \Rpole  & \vcrit & $\log{\Lstar}$ \\
Subtype  & (\Msun) & (\Rsun) & (\kms) & (dex \Lsun) \\ \hline
B0   & 17.5 & 7.7 & 538 & 4.64 \\
B0.5 & 14.6 & 6.9 & 519 & 4.41 \\
B1   & 12.5 & 6.3 & 502 & 4.21 \\
B1.5 & 10.8 & 5.7 & 491 & 4.01 \\
B2   &  9.6 & 5.4 & 475 & 3.85 \\
B2.5 &  8.6 & 5.0 & 468 & 3.68 \\     
B3   &  7.7 & 4.7 & 456 & 3.52 \\
B4   &  6.4 & 4.2 & 440 & 3.24 \\
B5   &  5.5 & 3.8 & 429 & 3.00 \\
B6   &  4.8 & 3.5 & 418 & 2.78 \\
B7   &  4.2 & 3.2 & 408 & 2.56 \\
B8   &  3.8 & 3.0 & 401 & 2.39 \\
B9   &  3.4 & 2.8 & 393 & 2.20 \\ \hline
\end{tabular}
\end{table}


\section{Spectroscopic Simulations} \label{sec:spect}

\subsection{Method}

For both sets of surface models, we synthesize profiles for
\heone~$\lambda$4471 and \mgtwo~$\lambda$4481\,\AA\ (including doublet
and forbidden components), the lines most commonly used by observers
to determine B-star \vsini\ values. The calculations use the \bruce\
code \citep{Tow1997b}, which is readily adapted from its original
intended application in modeling nonradial pulsation by simply setting
the pulsation amplitudes to zero. 

A large grid of precomputed conventional model spectra is required by
\bruce, to describe the angular and wavelength dependences of the
photospheric radiative intensity over the range of \teff, \logg\
values encountered between pole and equator. We construct such a grid
using the \synspec\ spectral-synthesis code by I. Hubeny and
T. Lanz. For the B0--B5 models, the underlying atmospheric structure
is calculated using the non-LTE \tlusty\ code
\citep{Hub1988,HubLan1995}, under the assumption that hydrogen and
helium are the only sources of line opacity; for the later subtypes,
atmospheric models are taken from the line-blanketed LTE grids
published by \citet{Kur1993}.

\begin{figure}
\epsffile{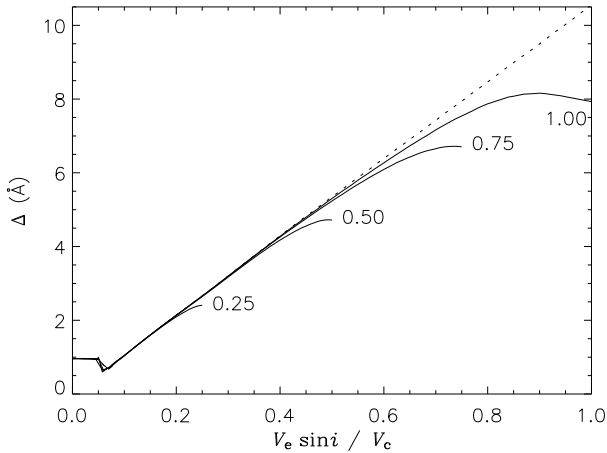}
\caption{The characteristic widths, \wid, of the \heone~$\lambda$4471\,\AA\
line profile for the B2 subtype, plotted as a function of \vesini\ for
the two models and four separate inclinations considered in the
text. The width data of the complete model (solid lines) are labeled
with their associated \sini\ values; those of the uniform model
(dotted line) all lie on the same curve. For $\vesini \lesssim
0.05\vcrit$, the intrinsic line broadening and blended components
dominate rotational effects, and the characteristic widths should be
disregarded.}
\label{fig:widths}
\end{figure}

The rotating-star spectral synthesis is undertaken over a range of
equatorial velocities $0<\ve<\vcrit$, and at four differing
inclinations: $\sini = \{0.25,0.50,0.75,1.00\}$. We characterize the
line widths of the resulting profiles by using the first zero,
$\xi_{1}$, of the Fourier transform of the flux spectrum
\citep{SmiGra1976}; specifically, we use $\wid \equiv \xi_{1}^{-1}$.
These characteristic widths \wid\ have a nearly linear relationship
with full-width at half depth, but in any case our results are largely
insensitive to the particular approach used to characterize line width.

To examine the robustness of our calculations, we computed a number of
variants on the basic models, including different choices of stellar
parameters \citep[e.g., those published by][]{Sle1992}, and
alternative parameterizations of the effects of rotation (e.g., fixing
the polar temperature, rather than varying it in accordance with
equation~\ref{eqn:tpole}). These variants introduce only small
quantitative changes in the results.

\subsection{Results: B2 subtype} \label{ssec:spect-b2}

\begin{figure*}
\epsffile{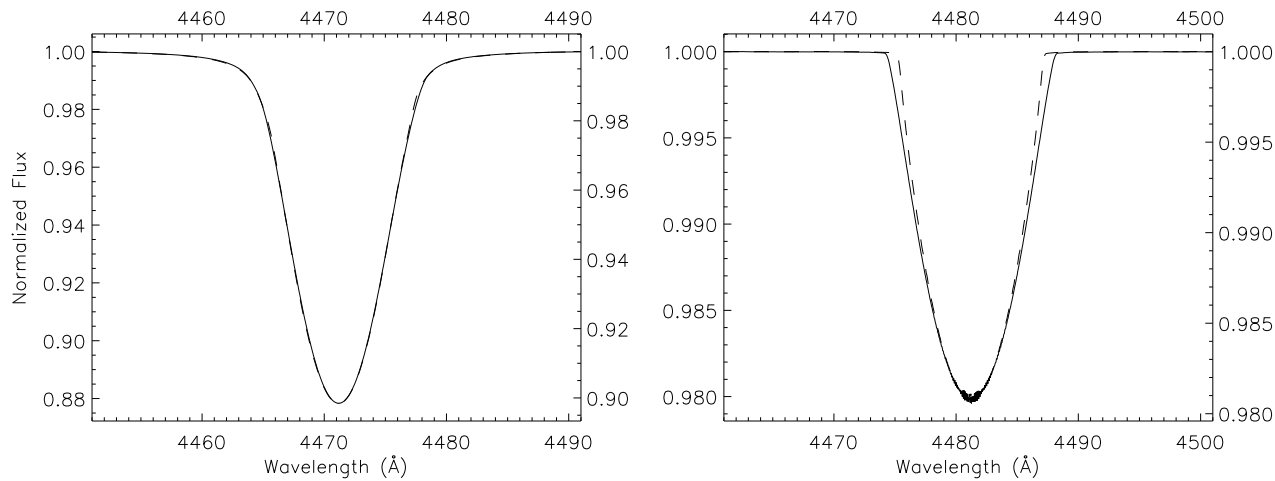}
\caption{\heone~$\lambda$4471 and \mgtwo~$\lambda$4481\,\AA\ line
profiles for equator-on `complete' B2 models at $\ve=395\,\kms$ (solid
lines, left-hand scales) and $\ve=460\,\kms$ (dashed lines, right-hand
scales). To facilitate the comparison of line widths and shapes, the
models are plotted on slightly different vertical scales, chosen so as
to normalize the line depths.}
\label{fig:profiles}
\end{figure*}

We first consider the surface models for the B2 subtype, the most
common in the sample of known Be stars \citep[see,
e.g.,][]{Por1996}. Fig.~\ref{fig:widths} shows the characteristic
widths of the \heone~$\lambda$4471\,\AA\ line as a function of
\vesini, at each of the four \sini\ values considered herein, for both
the uniform (\widu) and complete (\widc) models. We note that the four
solid lines each span a different \vesini\ range because the domain
$\ve = [0,\vcrit]$ has been scaled by the four different \sini\
values; a similar effect is not apparent in the dotted lines, because
the \widu\ curves of the uniform models are self-similar\footnote{By
this, we mean that one curve can be transformed into another by
scaling both \widu\ and \vesini\ by the same amount}, and lie atop one
another. The latter result is a corollary of the fact that line widths for the
uniform model follow a linear relationship with \vesini.


Unlike the uniform models, the characteristic width curves for the
gravity-darkened `complete' models (which furnish a better
representation of real stars) are \emph{not} self-similar, an
embodiment of departures from a linear \widc-\vesini\
relationship. Such departures arise from a number of effects,
including the variations of both line and continuum strength with
latitude, but the dominant factor for fast rotators is the decrease in
continuum emission near the equator resulting from von~Zeipel
darkening. This leads to a relatively small contribution from the
fast-rotating equatorial regions to the spatially integrated spectrum,
and in turn to a systematic reduction in line widths relative to the
uniform model. Thus if \vsini\ values are calculated using the
calibration found for the uniform model, they may underestimate the
true value of \vesini\ by $\sim$20\%, or more.  Moreover, the
near-invisibility of the equatorial regions in near-critical rotators
means that {\em increasing \ve\ has almost no effect on line width}
for such stars.

We illustrate this line-width redundancy in Fig.~\ref{fig:profiles},
for models with $\ve=395$ and $460\,\kms$ (0.83 and 0.97 \vcrit,
respectively).  Clearly, even with our effectively noise-free
synthetic data, the helium profiles are almost indistinguishable in
terms of their widths; there is no practicable means of ascertaining
that the two profiles shown belong to models with equatorial
velocities that differ by almost 20{\%}. This underlines the point
that the degeneracy of \widc\ as a function of \ve\ must seriously
compromise any attempt at devising a $\vsini(\wid)$ calibration, and
particularly a single-valued, near-linear calibration, for stars
strongly affected by gravity darkening.

Also shown in Fig.~\ref{fig:profiles} are the
\mgtwo~$\lambda$4481\,\AA\ profile for the same $\ve=395$ and
$460\,\kms$ models. The equivalent width (EW) of this line increases
with decreasing temperature (by a factor two from pole to equator in
our $\ve=0.95\vcrit$ models), unlike the \heone\ lines.  This increase
in relative line strength partly offsets the decline in continuum flux
(which falls by factor of nearly four, from pole to equator), and as a
result the Mg line widths show a somewhat greater (though still small)
response to increasing equatorial rotation.

\subsection{Results: other subtypes} \label{ssec:spect-other}

\begin{table}
\caption{Velocity deficiencies \vdef, as a percentage of the critical
velocity \vcrit, for the \heone~$\lambda$4471 and
\mgtwo~$\lambda$4481\,\AA\ profiles of the 13 model subtypes; in each
case, $\ve/\vcrit = 0.95$ and $i=90\degr$.}
\label{tab:vdef}
\begin{tabular}{@{}ccc} 
Sp.      & \multicolumn{2}{c}{$\vdef/\vcrit$ (\%)} \\
Subtype  & \heone~$\lambda$4471\,\AA & \mgtwo~$\lambda$4481\,\AA \\ \hline
B0   & 12.1 &  8.7 \\
B0.5 & 13.7 & 10.0 \\
B1   & 15.4 & 11.0 \\
B1.5 & 17.0 & 11.0 \\
B2   & 18.5 & 11.4 \\
B2.5 & 20.0 & 11.5 \\
B3   & 21.6 & 11.9 \\
B4   & 24.3 & 12.8 \\
B5   & 24.6 & 12.0 \\
B6   & 26.6 & 12.5 \\
B7   & 28.9 & 14.0 \\
B8   & 30.8 & 15.4 \\
B9   & 33.4 & 17.1 \\
\end{tabular}
\end{table}

The surface models for the other subtypes exhibit behaviour similar to
the B2 models discussed above: with the onset of gravity darkening,
the widths of line profiles saturate, and become insensitive to any
further increase in the equatorial velocity. To quantify succinctly the
degree of this effect, we measure the line width for complete models
at $\sin{i} = 1$, $\ve/\vcrit = 0.95$, and then use the uniform models
to infer a \vsini\ value for that width. We denote the difference
$(\vesini -\vsini)$ as a `velocity deficiency', \vdef, which we list
in Table~\ref{tab:vdef} for each of the 13 subtypes considered.

In both the \heone\ and \mgtwo\ lines, there is a systematic increase
in the velocity deficiency (expressed as a fraction of \vcrit) toward
later spectral types. For the most part, this increase arises from an
enhancement of the continuum gravity darkening effect, as measured by
the ratio between polar and equatorial continuum fluxes: in the
$\lambda\lambda$4471--4481\AA\, region appropriate to the \heone\ and
\mgtwo\ lines, this ratio is 3.7 for the B0 models, but grows to 7.0
for the B9 models. We note in passing that when \emph{bolometric}
fluxes are considered, the flux ratio is fixed at $\approx 14.7$ by
dint of the fact that $\ve/\vcrit = 0.95$, irrespective of which model
is under consideration.

The overall increase in $\vdef/\vcrit$ toward later spectral types is
modulated by the dependence of line EW on temperature. As discussed in
the preceding section, the EW variation of the \mgtwo\ line tends to
counteract the continuum gravity darkening effect. This explains why
the velocity deficiencies exhibited by the line are smaller than those
of the \heone\ line, and also why the growth in $\vdef/\vcrit$ is seen
to reverse itself temporarily around subtype B5.


\section{Photometric Simulations} \label{sec:phot}

\subsection{Method}

In the preceding section, we have demonstrated how gravity darkening
can make a near-critical star appear, when observed spectroscopically,
to be rotating at an ostensibly slower rate. In this section, we
broaden our investigation by examining the impact of gravity darkening
on the \emph{photometric} signatures of rapidly rotating stars.

For the complete surface models (\cf\ Section~\ref{sec:models}), we
synthesize absolute magnitudes in the Johnson $B$ and $V$ bands. The
calculations use our \rotcolor\ code, which is based on the
\bruce\ code but oriented toward photometric rather than spectroscopic
modeling. \rotcolor\ requires an input grid of precomputed
photometric fluxes, which we obtain by combining the four-coefficient
limb-darkening data published by \citet{Cla2000} with normal-emergent
absolute flux data made available to us by Claret (personal
communication); both data are based on the \citet{Kur1993} atmosphere
grids. To convert synthetic fluxes into absolute magnitudes, we use
the observed values for the Sun, $M_{B}=5.48$ and $M_{V}=4.83$
\citep{All1976}.

Our calculations are undertaken for five rotation velocities
$\ve/\vcrit = \{0.00,0.24,0.48,0.71,0.95\}$, and at three differing
inclinations $i = \{0\degr,45\degr,90\degr\}$. For comparison
purposes, we also synthesize colours for the evolutionary tracks
calculated in Section~\ref{sec:models}; each track extends from
zero-age main sequence (ZAMS) through to terminal-age main sequence
(TAMS, defined by the cessation of core hydrogen burning).

\subsection{Results}

\begin{figure}
\epsffile{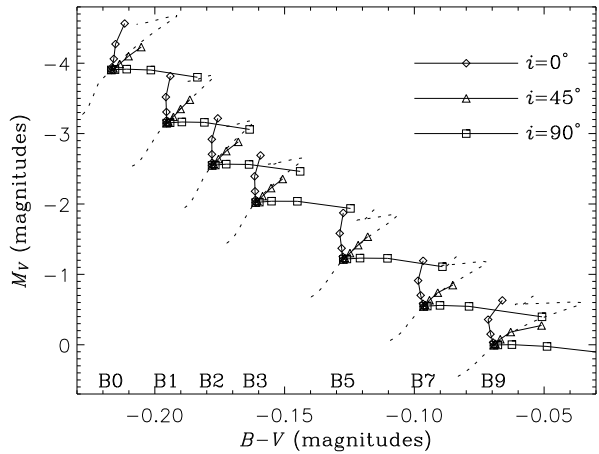}
\caption{Selected complete surface models (\cf\ Section~\ref{sec:models})
plotted as points in the Johnson $B-V$ colour-magnitude diagram, at
five rotation velocities $\ve/\vcrit = \{0.00,0.24,0.48,0.71,0.95\}$
and three inclinations $i = \{0\degr,45\degr,90\degr\}$. Solid lines
join together the points from models that differ solely in their
rotation velocity; dotted lines indicate ZAMS--TAMS evolutionary
tracks that pass through the non-rotating models, the latter being
labeled at the bottom of the plot with their nominal spectral types.}
\label{fig:color}
\end{figure}

In Fig.~\ref{fig:color} we plot the results of our photometric
simulations in a colour-magnitude diagram. The general effect of
rotation is to displace models away from their zero-rotation loci,
moving them along trajectories in the $M_{V}$/$(B-V)$ plane that are, to a
first approximation, straight lines. The differential rate of
displacement varies in step with \ve, being largest close to critical
rotation, and so small for the slowly rotating models
($\ve=0.24\,\vcrit$) that they can barely be distinguished from their
non-rotating neighbours.

The most striking aspect of Fig.~\ref{fig:color} is the strong
inclination ($i$) dependence of the trajectories' orientation in the
$M_{V}$/$(B-V)$ plane. For the $i=0\degr$ models, rotation increases
the observed brightness, due to beaming of radiation from the hot
pole; however, this increase happens without any appreciable variation
in colour, leading to trajectories that are close to vertical. The
converse occurs when $i=90\degr$: the centrifugal distention of
equatorial regions produces a drop in the surface-averaged effective
temperature, reddening the colours at approximately constant
brightness, and so resulting in nearly horizontal trajectories. For
intermediate $i=45\degr$ case, a combination of brightening and
reddening produces diagonal trajectories, coincidentally nearly
parallel to those produced by the process of stellar evolution.

The increase in brightness for the pole-on models is approximately
independent of subtype, being $\Delta M_{V} \approx -0\fm65$ at $\ve =
0.95\,\vcrit$, while the reddening of the equator-on models, at the
same rotation rate, shows a modest increase from $\Delta(B-V)=0\fm032$
at B0 to $\Delta(B-V)=0\fm066$ at B9. By chance, the ratios between
these brightening and reddening values correspond closely to the
colour-magnitude slope of the main sequence throughout the B spectral
type. Therefore, for an ensemble of randomly oriented B stars, the
effect of gravity darkening due to rotation at fixed $\ve/\vcrit$ is
to shift the main sequence up and to the right in the colour-magnitude
diagram, but to leave it otherwise undistorted. We will return to this
point in Section~\ref{ssec:disc-phot}.


\section{Discussion}

The calculations in Section~\ref{sec:spect} are intended to underline
our central point that line widths are a rather poor indicator of
\vesini\ at near-critical rotation.  Although this basic idea has been
previously noted in the literature (\citealt{Sto1968b, HarStr1968b,
ColTru1995}), in practice it has been almost entirely ignored by observers
\citep[cf.][]{How2003}. For example, in the course of writing up this work we
discovered that \citet[][ his Fig.~4]{Sto1968b} gives a diagram that
is very similar to our Fig.~\ref{fig:widths}, while also addressing
the implications for Be stars.  However, the low citation rate for
this paper (at the time of writing, a single
citation\footnote{\citet{ColTru1995} cite Stoeckley's work, and also
give a similar diagram to our Fig.~\ref{fig:widths} (their Fig.~4);
but we disagree with their conclusion that an approximate treatment of
limb darkening is the major factor in departures from the results of
the uniform model. Note that even our simple, uniform models
incorporate a detailed treatment of non-linear, wavelength-dependent
limb darkening.}  in the ADS database since 1975) suggests that the
significance of this result for Be-star physics has been largely
overlooked.
In this section, therefore, we critically review 
the historical basis
for the canonical view that \ve/\vcrit\ does not
exceed $\sim$0.8 for Be stars.

\subsection{Line profiles}

The basic method for determinations of \vsini\ is, of course,
line-profile measurements.  Slettebak's (1982) observational study of
rotation in a sample of 163 Be stars has been particularly
influential; many authors have relied directly on his \vsini\ data
\citep[\eg,][]{Por1996}, or have used them to calibrate
\citep[\eg,][]{Bal1975,Hal1996,Ste1999,Yud2001,Abt2002} or to validate
\citep[\eg,][]{How1997,Cha2001} their own measurements.

\citet{Sle1982} used visual inspection of photographic spectra to
estimate the line widths in his targets, calibrating his results with
standards from \citet{Sle1975}.  This approach assumes a single-valued
correspondence between line width and \vsini, and neglects the
redundancy between these parameters that we have emphasized.
Slettebak is universally acknowledged as an exceptionally skilled
observer, and we believe that his careful and conservative approach
would have led him to adopt the {\em minimum} \vsini\ value consistent
with his observations (rather than some arbitrarily larger value),
which, as we have shown, will generally underestimate \vesini.

Furthermore, the \vsini\ values of the underpinning \citet{Sle1975}
standard stars were themselves calibrated against theoretical profiles
from \citet{Col1974}. The latter's pioneering calculations required
computations that were, by the standards of the day, sophisticated and
demanding.  Unfortunately, subsequent results, including his own
\citep[\cf\,][]{ColTru1995} and those we present here, shed some doubt on this
fundamental calibration.  For example, \citet{Col1974} found line
broadening to be relatively unaffected by gravity darkening. Indeed,
his paper explicitly notes a ``remarkably good linear correlation
between the theoretical value for \vesini\ and the half-width of the
line'' -- a correlation from which our gravity-darkened, complete
models depart, quite significantly, in the crucial limit of large
\vesini\ (Fig.~\ref{fig:widths}).

Collins' remark was illustrated by calculations of the
\heone~$\lambda$4471\,\AA\ profile for a B0-type stellar model (his Fig.~5).
That figure indicates line half-widths at half depth that are
systematically \emph{broader} for critical rotators than for
sub-critical rotators at the same \vesini.  We find it difficult to
devise any physical explanation for such behaviour, and suspect that
it may indicate some procedural or coding error in his study. Indeed,
later work by, e.g., \citet{ColTru1995} gives findings broadly
consistent with our own, and also contradicts the linear correlation
found by \citet{Col1974}.  In any event, our Table~\ref{tab:vdef}
indicates that the B0 subtype exhibits the smallest velocity
deficiencies of any of the stellar models that we have
considered. Therefore, notwithstanding any putative problems with
Collins' early calculations, we expect the data plotted in his Fig.~5
to provide the poorest illustration of the degree to which gravity
darkening can influence line broadening.

From the point of view of both observational procedure and underlying
calibration, we are therefore led to the conjecture that the system of
Be-star velocities underpinned by the work of \citet{Sle1975} and
\citet{Sle1982} may systematically underestimate \vesini\ values,
particularly at the upper end of the rotation scale.  This conjecture
is given credence by the recent investigation by \citet{Cha2001}, who
obtained \vsini\ values for 116 Be stars by least-squares fits of
synthetic spectra to observed \heone~$\lambda$4471\,\AA\ line profiles.  As
for our uniform models, Chauville et al.'s synthetic profiles were
calculated under the assumption of no rotational distortion or gravity
darkening, and they therefore certainly underestimate \vesini.
However, they obtained \vsini\ values in excellent overall agreement
with those of \citet{Sle1982}, encouraging the view that there is also
a systematic error in the latter's results (and hence in all
subsequent work built on them).

\subsection{Statistics}

\citet{Cha2001} measured a mean projected equatorial velocity
$\langle\vsini/\vcrit\rangle = 0.65$ for their sample of Be stars. Using
the \citet{ChaMun1950} relations between projected and intrinsic rotation
velocities, they inferred from their data a mean equatorial velocity
$\langle\ve/\vcrit\rangle = 0.83\pm0.03$.
This extremely small apparent
intrinsic scatter led them to conclude that ``all studied Be stars
rotate at nearly the same ratio [of] $\ve/\vcrit$''.

We know of no physical reason why all Be stars should rotate at the
same $\ve/\vcrit \simeq 0.8$, but examination of our
Fig.~\ref{fig:widths} shows that at and above $\ve/\vcrit \simeq 0.8$
there is effectively no change in line width with increasing \ve. It
is therefore reasonable to suppose that the observations may be
consistent with {\em any} value (or a range of values) $\gtrsim0.8$,
and particularly with near-critical rotation -- an obvious limiting
value.

To investigate this possibility in a heuristic way, we consider a
population of complete B2-star models with equatorial velocities
distributed uniformly between $\ve=0.93\,\vcrit$ and
$\ve=0.97\,\vcrit$, and with rotation axes oriented
randomly. Analyzing this population with \heone~$\lambda$4471\,\AA\ line
profiles derived from our uniform models (thereby effectively
mimicking Chauville et al.'s procedure), we obtain a mean projected
equatorial velocity $\langle\vsini/\vcrit\rangle = 0.635$, from which the
\citet{ChaMun1950} relations suggest a mean equatorial velocity
$\langle\ve/\vcrit\rangle = 0.809$, with a variance
$\sigma^{2}(\ve/\vcrit) = -9\times10^{-3}$.  These \emph{inferred}
values\footnote{Note that the variance for our synthetic population is
less than zero due to the marginal breakdown of the assumption by
\citet{ChaMun1950} that the \vsini\ distribution arises from the
spatial projection of an intrinsic distribution.} are remarkably close
to those obtained by Chauville et al., even though our synthetic
population rotates at significantly more-rapid rates than the
$\langle\ve/\vcrit\rangle = 0.83$ advanced by those authors. Of course,
this simulation is not particularly realistic, but it nonetheless
clearly establishes that statistical studies that do not adequately
treat the effects of gravity darkening must inevitably lead to
systematic underestimates of \vesini.

\subsection{Photometry} \label{ssec:disc-phot}

The calculations in Section~\ref{sec:phot} indicate that near-critical
rotation (by which we mean $\ve/\vcrit \simeq 0.95$) displaces stars
toward brighter magnitudes or redder colours, depending on the
inclination. This phenomenon, first discovered by \citet{RoxStr1965},
has been investigated and confirmed by a number of authors
\citep[\eg,][]{Col1966,Fau1968,HarStr1968a,MaePey1970}. Our
qualitative findings are very similar to these studies, although we
obtain smaller changes in brightness ($\Delta M_{V} \simeq -0\fm6$)
and colour ($\Delta(B-V) \simeq 0\fm06$) than found historically (for
reasons we discuss below).

Evidently, if Be stars are indeed near-critical rotators, they should
naturally occupy a brighter/redder position in the colour-magnitude
diagram than normal B-type stars. A search through the relevant
literature reveals substantive evidence for such an anomalous
position: \citet{ZorBri1991} find their sample of Be stars to be
over-luminous with respect to main sequence stars, and of the fourteen
previous studies that they cite, eleven disclose similar behaviour.

In the era when the observational consequences of gravity darkening
were first being investigated, it was natural to attribute this
observed over-luminosity to rapid rotation
\citep[\eg,][]{Sto1968b}. However, many of the studies from the 1960's
period (including the majority of those we cite above) were based on
interior calculations by \citet{Rox1965}, which were found by
\citet{San1970} to contain a serious error. Correction of this error
resulted in a reduction of the brightness/reddening change caused by
rotation (towards values commensurate with those we found in
Section~\ref{sec:phot}), to such an extent that rotation alone was
hard-pushed to explain the luminosity anomaly.

Ever since the discovery of the error by \citet{Rox1965}, there has
been a shift towards interpreting the anomaly as a consequence of the
circumstellar emission (CE) produced by the disks of Be stars
\citep[\eg,][]{Fab1996}. This hypothesis certainly seems promising,
but it has never been established whether CE can explain the
observations \emph{in toto}; as \citet{ZorBri1997} point out, it may
be the case that multiple mechanisms (including rapid rotation) are
required to explain the anomalous position of Be stars in the
colour-magnitude diagram. Indeed, it is still not clear what the
effects are of the intrinsic `rotation reddening' discussed by
\citet{Mae1975}, which arise from the distention of the envelope
around the equatorial regions of a rapidly rotating star (such that
the local atmospheric structure departs from that predicted by
plane-parallel models).


\subsection{Future Approaches} \label{ssec:future}

Having presented evidence that current measurements of Be star
rotation are underestimates, let us briefly consider what approaches
might be deployed in the future to address this problem. To reiterate,
the essence of the problem is this: when a star rotates at a
significant fraction of its critical velocity \vcrit, the
insensitivity of line widths toward any further increase in the
rotation rate means that these widths are poor indicators of the
star's true projected equatorial velocity \vesini. How might this
degeneracy be lifted?

One possible approach lies in leveraging the combined
temperature/wavelength dependence of the continuum flux. From far
ultraviolet (FUV) observations of Be stars, \citet{Hea1976} discovered
the line broadening at these wavelengths to be significantly less than
at visible wavelengths. As pointed out by \citet{Hut1976}, this effect
can be understood as a result of the different surface brightness
distributions at visible and FUV wavelengths, caused by gravity
darkening. \citet{HutSto1977} subsequently suggested that the effect
might be used to obtain reliable values for $\ve/\vcrit$ and $i$. Such
an approach certainly looks promising, but suffers from the drawback
of requiring simultaneous visible, FUV and (if possible) infrared
observations, which are often difficult to secure.

An alternative technique for lifting the degeneracy lies in exploiting
the differing temperature sensitivities of different spectral
lines. As we demonstrated in
Sections~\ref{ssec:spect-b2}--\ref{ssec:spect-other}, the inverse
temperature/EW dependence exhibited by the \mgtwo~$\lambda$4481\,\AA\
line partly counteracts the effect of gravity darkening in the
continuum, and lessens the impact of the degeneracy. In the case of an
individual line such as this, the effect will probably be too small to
be of practical use\footnote{We also note on more general grounds that
the \mgtwo~$\lambda$4481\,\AA\ line is a poor choice for \vesini\
determination, due to blending with the adjacent
\heone~$\lambda$4471\,\AA\ line.}, but the effect could be exploited
by simultaneous modelling of \emph{multiple} lines. In principle, the
different temperature/EW dependence of each line means that a good fit
for \emph{all} lines will occur only at a single point in parameter
space (that is, a single combination of \ve, $i$, \Mstar\ \etc). In
reality, of course, noise and other factors will limit the precision
of this approach; nevertheless, the approach certainly appears
promising, and has already been applied with success to
rapidly-rotating O stars \citep[see][]{HowSmi2001}.

The most direct method for examining rotational distortion is through
direct imaging. Once the aspect ratio $x \equiv \Req/\Rpole$ of the
centrifugally-distorted star is known, its equatorial velocity can be
calculated in the Roche approximation from the expression
\begin{equation}
\frac{\ve}{\vcrit} = x \sqrt{\frac{3(x-1)}{x^{3}}},
\end{equation}
where we note that $x=1.5$ corresponds to $\ve/\vcrit=1$. Recently,
\citet{deSou2003} applied this approach to the Be star $\alpha$ Eri;
their \emph{VLTI} observations indicate an aspect ratio entirely
consistent with near-critical rotation. Clearly, the power of this
approach lies in its ability to avoid the need for complex modelling
of stellar atmospheres or interiors. Unfortunately, its applicability
is limited to those stars which are resolvable via interferometry.

\section{Conclusion}

We have argued that the available observations, and their
interpretation, leave open the possibility that Be stars rotate at or
near to their critical velocities, contrary to the previous consensus.
In questioning the canonical view that Be-star rotation is
substantially sub-critical, we have been motivated by the issue of the
fundamental physical processes responsible for the Be phenomenon.  A
range of phenomenological models have been developed to account for
various aspects of Be-star behaviour, some quite detailed and
sophisticated (e.g., \citealt{Oka1991}), but the mechanisms underlying
the formation of circumstellar disks remain elusive.  This is largely
due to the substantial energetic requirements for levitating material
in a strong gravitational field.

Near-critical rotation alleviates this problem to the point where
known processes, such as pulsation or gas pressure, may readily
provide sufficient energy and angular momentum. Our critique argues
that such near-critical rotation \emph{has not} been ruled out by
existing studies, be they spectroscopic \emph{or} photometric. Indeed,
these studies furnish \emph{necessary} evidence that the Be stars
\emph{are} rotating close to critical; we hope that this paper will
provide a motivation towards obtaining, perhaps using one of the
approaches we suggest in Section~\ref{ssec:future}, \emph{sufficient}
evidence in support of the same conclusion.


\section*{Acknowledgments}
We thank Dietrich Baade and Thomas Rivinius for their helpful
suggestions toward improving this paper.  RHDT acknowledges PPARC
support.  SPO received support from a PPARC fellowship for sabbatical
research in the UK, and from NSF grant AST00-97983 to the University
of Delaware. IDH is a Jolligoode fellow.


\bibliography{paper}

\bibliographystyle{mn2e}


\label{lastpage}

\end{document}